# Analyze Drivers' Intervention Behavior During Autonomous Driving – A VR-incorporated Approach

Zheng Xu

*Department of Civil Engineering*
*Monash University*
*Clayton, VIC 3800, Australia*

Zheng.Xu3@monash.edu

*Abstract* – Given the rapid advance in ITS technologies, future mobility is pointing to vehicular autonomy. However, there is still a long way before full automation, and human intervention is required. This work sheds light on understanding human drivers' intervention behavior involved in the operation of autonomous vehicles (AVs) and utilizes this knowledge to improve the perception of critical driving scenarios. Experiment environments were implemented where the virtual reality (VR) and traffic micro-simulation are integrated, and tests were carried out under typical and diverse traffic scenes. Performance indicators such as the probability of intervention, accident rates are defined and used to quantify and compare the risk levels. By offering novel insights into drivers' intervention behavior, this work will help improve the performances of the automated control under similar scenarios. Furthermore, such an integrated and immersive tool for autonomous driving studies will be valuable for research on human-to-automation trust. To the best knowledge of the authors, this work is among the pioneer works making efforts into such types of tools.

## I. INTRODUCTION

The introduction of autonomous vehicles (AVs) brings a significant breakthrough into the transportation field, and it will reshape the mobility system, e.g., automation-supported, safer, and faster [1]. It is a common recognition that full automation (autonomous driving at Level 5) is still far from reality, because vehicles would rely on both advanced vehicular control systems and environmental sensor systems that require time for development and specific infrastructure for operation [2]. Therefore, the near-future mobility system is envisioned to consist of mixed or semi-automated (correspond to Levels 3-4) traffic flow.

From AV deployment perspective, over a billion-kilometers level of tests are typically required to verify the functional reliability of AVs [3], which is challenging on both spatial and financial sides. Furthermore, the public acceptance of AVs is low, as most people are skeptical of the safety and usefulness of AVs [1]-[2], which again is challenging to tackle given the lack of testing opportunities. It is important to understand the features of mixed traffic behavior, particularly because the interaction between humans and AVs is essential to safety. It appears to be difficult to rely on the existing implementation cycle to obtain useful feedback and knowledge [4].

From AV knowledge perspective, existing literature focused on the performance of AVs when facing certain interruptive conditions during operations [5]-[6], and suggested human input is crucial to avoid unexpected hazards. Obviously, this is a trust issue of human-to-automation and not consistent with the official definition of Level 4 and 5 autonomous driving. For instance, studies concluded that human-to-machine trust is compassionate under high-risk circumstances such as highway driving [5] and dilemma situations [6]. Fundamentally, this reflects the insufficiency of the current control of the AVs, as they cannot completely identify risks nor react to risks accurately in certain scenarios.

From AV research toll perspective, many VR supported works exist in literature. Nevertheless, most studies focused on offering a user experience of sitting on AVs [7] or emphasized pedestrian-AV interaction [8], rather than studying the AVs and behaviors of any human interactions from drivers' perspectives.

This work aims to gain insights on the perceptions and reactions of drivers when intervention may be required during autonomous driving; specifically, this work will: 1). develop AVs in a VR and micro-simulation integrated platform; 2). define and identify the critical driving scenarios in which drivers tend to perform driving intervention behavior (DIB); 3). demonstrate the DIB under various and disruptive scenarios.

## II. METHOD

### A. AV Development in the Simulation Environment

The AVs were developed in a validated simulation platform [9]-[10] with the assistance of the Artificial Neural Networks (ANN) and Genetic Algorithms in the form of a client-server model. The developing environment contains both urban driving and freeway merging sections. Traffic flow simulated by VISSIM could be seen and interacted with AVs via the VR interface. Combining the strengthen of traffic reproduction by calibrated micro-simulation and the immersive and high-fidelity environment by VR, the platform is suitable for AV-related development and testing.

The training dataset consisted of 500 vehicle trajectories generated from VISSIM and 46 human-conducted drives extracted based on the covariates obtained from the experiment subjects in [9]-[10]. As the training is about mapping the input, such as distance between obstacles and vehicle speed, to control actions such as acceleration and steering angle, the ANN is applied for learning the mapping structure of the given data. Then GA is applied to optimize the control output from discrete values to continuous space. As such, the calibrated AV models can form their intelligence, e.g., completely risk-free while driving. Fig.1 illustrates the key features in this study.

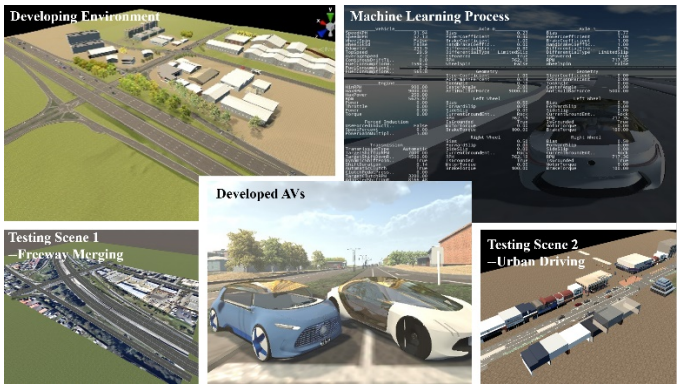

Fig. 1 Key features in this study.

## B. Experiment Design

All the scenes are real-world replications, e.g., the urban driving scene is a reproduction of a local area in Melbourne, Australia, consisting of a typical arterial road crossing in a local business center with a speed limit of 50km/h. The freeway merging sections come from Melbourne real cases, reflecting two common geometric designs for on-ramps (straight and bend roads with 15-degree slopes). Meanwhile, these sites are critical "black spots" according to Victoria Crash Statistics [11]. A critical scenario may have one or several of the following characteristics: merging difficulty or significant speed reduction due to high traffic volume (> 2500veh/h on urban arterials and > 8500veh/h on freeways), significant speed reduction or vision deprivation due to adverse weather conditions (heavy rain, snow, and fog), and unexpected interactions from road users of the conflicting flow.

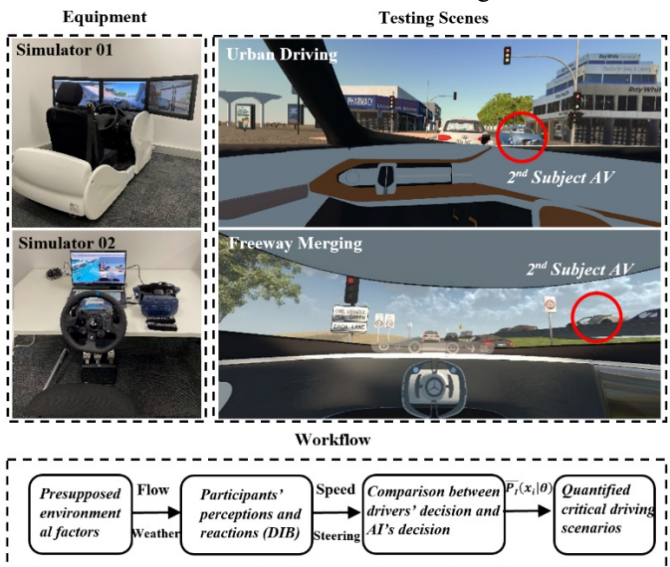

Fig. 2 The framework of this study.

Six participants (4 males, 2 females, mean = 27.33, SD = 3.45) were recruited for the experiment. They were instructed to intervene during the AV operation if they feel necessary. In this work, two driving simulators were utilized: a SIMREX CDS manufactured by Innosimulation of Seoul, and a Logitech G923 driving wheel and pedals. HTC VIVE Pro Eye was the VR interface for the participants, and it also functions as eye-tracking data collection.

Every two participants were formed in pair and put into one scene. In the freeway merging cases, one participant would be positioned on the on-ramp and the other would be on the mainstream; while for urban scene, the two participants would be positioned on the same lane or opposite lanes. As such all perspectives are covered. A total of 48 experiments were conducted. Questionnaires were distributed to the participants for information collection, and the following information were collected and recorded during the experiment: participants' responses, heart rates, and gaze patterns.

## C. Critical Scenario Analysis

Firstly, a scenario that contains higher conflict frequency $P(C)$ is logically defined as critical. Given that the presupposed critical scenarios might not be genuinely perceived as actual critical by participants or AVs, numerical assistance is required to quantify such scenarios based on the user's performance. This can also clarify testing scenarios and avoid missing scenarios that could not be generally defined as critical, resulting in a balance between exploitation and exploration in critical scenario determination.

Referring to the modified greedy sampling policy that suits for defining critical scenarios for connected autonomous vehicles (CAV) proposed by Feng et al. (2021) [12], the probability of drivers' decisions under such circumstances can be therefore quantified. As shown in the equations below:

- $\epsilon$-greedy Sampling Policy

$$\bar{P}_I(x_i|\theta) = \begin{cases} \frac{(1-\epsilon)V(x_i|\theta)}{W} & (x_i \in \varphi) \\ \frac{\epsilon}{N(X)-N(\varphi)} & (x_i \in X \backslash \varphi) \end{cases} \quad (1)$$

In the equations, $W = \sum_{x_i \in \varphi} V(x_i|\theta)$ indicating a normalization factor; $x_i$ denotes decision variables of testing scenarios; $X$ denotes feasible set of decision variables; $\varphi$ represents critical scenarios. $\bar{P}_I(x_i|\theta)$ measures the probability that the AV encounters the interaction event due to critical scenarios. The $\epsilon$-greedy sampling represents a small probability of the scenarios that beyond the normal definitions (presupposed scenarios). This method is commonly utilized for balancing exploitation and exploration.

Finally, the scenario criticality is represented by the comprehensive probability considering both $P(C)$ and $\bar{P}_I(x_i|\theta)$.

## III. RESULTS

### A. Performance of the Developed AV

Fig. 3a shows the evaluation results of the developed AV model in relation to accident rates. Performance indicators also include space-mean speed and tractor compliance, and they formed the same trend. There was a 10% to 20% decrease in vehicle velocity corresponding to different weather conditions based on the vehicle traction. It can be observed that with sufficient training runs (after 40 simulations), the performance of the AVs becomes stable.

## B. Experiment Results
### 1) Participants Reactions

A significant interaction between DIB and fog conditions was identified (p<0.005). Most DIB occurred during the fog condition in both scenes. Cut-in and pedestrian crossing were the causes of DIB in the urban driving scene; DIB occurred when it was difficult to merge due to the narrowed vehicle headways in the freeway merging scene. Drivers tended to slow down or stop before the AV reacted. Participants were more likely to perform DIB during fine weather conditions than rain and snow. Fig.3b shows the likelihood of DIB in different scenes under various conditions.

Traffic flow significantly affected drivers' willingness to perform DIB (p < 0.005). All the interventions appeared when traffic volume was high.

AV's traveling speed also significantly affected participants' reaction (p<0.005), reflecting as the deceleration behaviors while the AVs interacted with other road users. Fig.3c demonstrates the correlation between AV's relative speed, driving distance, and the number of interventions. Table 1 summarizes the results of the experiment.

TABLE I
SUMMARY OF THE EXPERIMENT RESULTS

| Typical Critical Scenarios | | Criticality | AV's Decision | Human subjects 1 | | Human subjects 2 | |
|---|---|---|---|---|---|---|---|
| | | | | DIB | AHR | DIB | AHR |
| UD | Cut-in | 0.72 | D | T&D | 98 | T&D | 97 |
| | Ped. | 0.71 | S | S | 102 | S | 100 |
| FM | Merging | 0.68 | A | D | 105 | N | 88 |
| | Cut-in | 0.65 | D | N | 97 | N | 85 |

*UD: urban driving; FM: freeway merging; Ped.: pedestrian; D: deceleration; S: stop; T: turning; N: none. All indicators are statistically significant (p < 0.005)

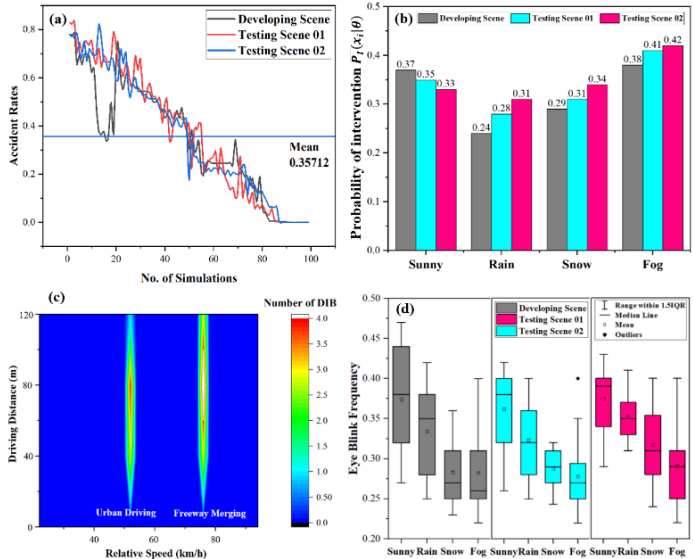

Fig. 3 Summary of the results

### 2) Participants Perceptions

Going deeper into the causes of such intervention behavior, the eye blink frequency and average heart rate (AHR) demonstrated that participants were more concentrated and excited while AVs were moving at a relatively high speed (> 70km/h) or in a relatively crowded area. With the eye blink frequency baseline of 0.4 blink/sec, participants blinked less during the adverse weather conditions in all the scenes. Fig. 3d illustrates the distribution of participants' eye blink frequency during the experiment. Based on their 3-minute AHR baseline, a minimum 15 increase was seen during the experiment, indicating a relatively high driving workload.

## IV. DISCUSSION

Interestingly, despite the high concentration during autonomous driving under rain and snow conditions, the probability of interventions was lower compared to the fine weather condition. This was principally due to the impact of moving speed on people's perception. In this study, although the AV was validated to maintain the accident rate at a deficient level (verge to 0), participants still worried about safety and felt uncomfortable while they could not take control during the critical scenarios. On the other hand, results illustrated the difference between the perceived hazards of humans and AI. Based on the proposed standard of the scenario criticality, 0.65 is proved to be the threshold that human drivers would perform DIB (e.g., deceleration or stop). Contrastingly, scenarios in which AVs tended to decelerate have the criticality of at least 0.72.

The significance of this work stems from both the research and practical aspects. The development of AVs in integrated simulation platform is a validated approach to guarantee the authenticity of the AVs. The VR experiment, models, and results of this study can be of use to manufacturers of AVs or researchers seeking to implement a more flexible, risk free, and high-fidelity approach in their work.

SUPPLEMENTARY MATERIALS

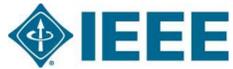

*IEEE Intelligent Transportation Systems Society*

The 2021 "Shape the Future of ITS" Competition:
Young Professionals Category – Third Prize

is presented to

*"Analyze Drivers' Intervention Behavior During Autonomous Driving – A VR-incorporated Approach"*

Zheng Xu
Monash University, Australia

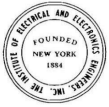
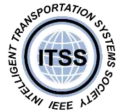

*Professor Emeritus Ljubo Vlacic*
*Chair, The Competition Evaluation Committee*
*Editor-in-Chief*
*IEEE Intelligent Transportation Systems Magazine*